# Visualizing the in-gap states in domain boundaries of ultra-thin topological insulator films

Jun Zhang[1], Junbo Chen[1], Shuaihua Ji[2], Yeping Jiang[1]*

[1] Key Laboratory of Polar Materials and Devices, Department of Electronic, School of Physics and Electronic Science, East China Normal University, Shanghai 200241, People's Republic of China

[2] State Key Laboratory of Low-Dimensional Quantum Physics, Department of Physics, Tsinghua University, Beijing 100084, People's Republic of China

Ultra-thin topological insulators provide a platform of realizing many exotic phenomena such a Quantum spin Hall effect, Quantum anomalous Hall effect, etc. These effects or states are characterized by quantized transport behavior of edge states. Experimentally, although these states have been realized in various systems, the temperature for the edge states to be the dominating channel in transport is extremely low, contrary to the fact that the bulk gap is usually in the order of a few tens of milli-electron volts. There must be other in-gap conduction channels that won't freeze out till a much low temperature. Here we grow ultra-thin topological insulator $Bi_2Te_3$ and $Sb_2Te_3$ films by molecular beam epitaxy and investigate the structures of domain boundaries in these films. By scanning tunneling microscopy and spectroscopy we find that the domain boundaries with large rotation angles have pronounced in-gap bound states, through which one dimensional conduction channels are suggested to form as visualized by spatially resolved spectroscopy. Our work indicates the critical role played by domain boundaries in degrading the transport properties.

In recent years, the concept of band topology has led to a rich family of topological materials. Among these materials, the Quantum spin Hall (QSH) states and Quantum anomalous Hall (QAH) states are mostly pursued by the research community. These systems are characterized by the dissipationless one dimensional (1D) conduction channels at the edge. Experimentally, although there are many proposed QSH and QAH materials, the quantized edge transport has only been realized in a few materials[1-9]. In addition, the temperature for the edge channels to be dominating is extremely low (< 1 K), except for the latest reported QSH transport in cleaved $WTe_2$ monolayer flakes (~ 100 K)[6]. For possible applications, the growth of ultra-thin films of QSH and QAH materials with relatively high working temperatures becomes a critical issue.

The two dimensional (2D) QSH and QAH materials is signified by the presence of an inverted bulk gap, which insures the existence of one dimensional (1D) edge states. The bulk gap normally is in the order of a few tens of milli-electron volts[10], suggesting a much higher working temperature compared with the experimental results. Thus, in the thin films of QSH and QAH materials there must be some limiting factors that prevent the edge channels to be dominating in transport. Molecular beam epitaxy is a technique to grow films with finely controlled thicknesses, especially ultra-thin films. Most proposed QSH and QAH materials are in the ultra-thin-film forms about a few nanometers. For such thin films grown on the substrates, various parameters such as the lattice mismatch, the growth dynamics, etc., will inevitably lead to the formation of various kinds of structural defects [11-16]. In this work, we mainly focus on the domain boundary structures. The domain boundaries, unlike other defects, cannot be easily removed by changing growth parameters. In addition, as suggested by the data here, the domain boundaries may introduce 1D conduction channels into the films.

The family of three dimensional (3D) strong topological insulators (TI) ($Bi_2Se_3$, $Bi_2Te_3$ and $Sb_2Te_3$) are characterized by an inverted bulk gap and a single Dirac-cone like surface state[17-20]. Various approaches have been proposed to get QSH and QAH states based on the materials in this family or their derivatives. Theoretically, the gap in ultra-thin 3D TIs can be tuned by thickness, electric field or exchange field into the inverted region to realize possible QSH or QAH states[1,21-23]. Experimentally, by introducing magnetic moments into the ultra-thin 3D TI films, QAH states have been realized in this family with low working temperatures about 1 K[7,8].

In addition, although the growth of high-quality TI films in this family is well developed and now routine[24-26], the investigation of film structures in the very vicinity of substrates is rare. By the growth of ultra-thin films and by a close investigation of the structures therein, we may find the key to improve the working temperatures of current QAH or QSH states, or even to realize the other proposed QSH or QAH systems. Thus, we focus on ultra-thin films of 3D TIs. Here we grow ultra-thin topological insulator $Bi_2Te_3$ and $Sb_2Te_3$ films and investigate the domain boundary structures and spatially dependent local density of states (DOS) of these structures. We find two dominant types of domain boundaries and they behave differently in DOS. One of them is suggested to be a possible factor that may provide conduction channels in the electronic transport.

We grow 2 quintuple layer (QL) $Bi_2Te_3$ and $Sb_2Te_3$ films on $SrTiO_3$ (111) (STO) substrates by molecular beam epitaxy (MBE). The substrates are annealed in vacuum prior to the deposition. The surface peak-to-peak roughness is about 200 pA. Figure 1a shows the topographic images for the 2-QL $Bi_2Te_3$ and $Sb_2Te_3$ films by scanning tunneling microscopy (STM). The films have dense domain structures probably because of the presence of strain in the films. The surface lattice constants of STO(111), $Bi_2Te_3$ and $Sb_2Te_3$ are ~ 3.90 Å, 4.38 Å and 4.26 Å, respectively. A close investigation of these films indicates that there are two kinds of domain boundaries marked as DW1 and DW2. These domain boundaries are formed because of the different crystal orientations of the adjacent domains. DW1 is dash-dotted like and the corresponding two domains have a relatively small rotation angle between them. Figure 1b shows a domain boundary of type DW1 with a rotation angle α of about 9°. Atomic resolution images (figure 1b and figure 2a) show that DW1 is characterized by the insert of an additional atom for every n lattice sites[11]. On the contrary, DW2 forms between domains with relatively large mismatch angles in crystal orientations. The lattices don't match along the boundary and the lattice seems strongly distorted. We checked domain boundaries' appearance in 2-QL $Bi_2Te_3$ and $Sb_2Te_3$ films. We find that the domain boundaries in 2-QL $Bi_2Te_3$ films are exclusively of type DW1. In contrast, DW1 and DW2 appear almost evenly in 2-QL $Sb_2Te_3$ films. A close investigation reveals that the two domains along DW2 are mainly have a rotation angle α of about 30° or about 60°. In the latter case the $\overline{\Gamma} - \overline{M}$ or $\overline{\Gamma} - \overline{K}$ directions of domains are opposite to each other as shown in figure 1c.

For DW1, we can relate the period of the dash-dotted like structure with the rotation angle α and do statistics on α for DW1 in both 2-QL $Bi_2Te_3$ and $Sb_2Te_3$ films. The results are plotted in figure 2b and 2c, respectively. Besides the fact that almost half of the domain boundaries are of type DW2 in $Sb_2Te_3$, the distribution of α for DW1 in $Sb_2Te_3$ also shifts to higher values by 2° compared with that in $Bi_2Te_3$. The relatively large α and the presence of DW2 in $Sb_2Te_3$ might indicate a weaker interaction between $Sb_2Te_3$ and STO. The possible stronger film-substrate interaction may help to force the $Bi_2Te_3$ domains to align with the crystal orientations in the substrate, while in $Sb_2Te_3$ the initial nucleation may tend to be randomly orientated. In the latter case, there appears an equal distribution between DW1 and DW2 for the domain boundaries in $Sb_2Te_3$.

By spatially resolved STS we find that DW1 and DW2 behave differently in disturbing the original electronic structure of materials. In the 2-QL $Bi_2Te_3$ film, spatially dependent STS are measured along the line sketched in figure 1b and plotted in figure 3b. Figure 3a shows typical STS for three different positions along DW1. The positions are also indicated by the black dots labeled by the numbers 1, 2 and 3, corresponding to the positions between the bright dots, about 0.5 nm away the dots and on the dots, respectively. We can see the formation of bound states periodically along DW1 on positions 1 and 3 at different energies (~ -60 meV and -290 meV). These bound states are bounded and separated from each other.

For the 2-QL $Sb_2Te_3$ film, we here show in figure 4a STS mapping at different energies on a uniform 2-QL region having DW1, DW2 and a step coming from the substrate. Figure 4b, 4c and 4d are the STS away from domain boundaries, on a bright spot in DW1 and on DW2, respectively. The gap shown in figure 4b is the hybridization gap for the 2-QL $Sb_2Te_3$ film[18]. The magnitude of gap is about 210 meV (approximately from -30 to -240 meV) and is uniform across the regions away from domain boundaries. On the bright spot in DW1 and on DW2, we find bound states starting from -170 meV and -60 meV, respectively. By taking STS mapping at energies of -150, -180 and -193 meV (figure 4a), we show here that the bound states in DW1 are localized near the bright spots along DW1, while those in DW2 are delocalized along DW2. All these three representative energies are within the hybridization gap and also within the energy range of bound states of DW2. -150 meV is within energy range of DW2's bound states but outside that of DW1s'. -180 and -193 meV are within the energy ranges both of DW1's and DW2's bound states. We can see that for all these three energies, the bound states are delocalized along DW2. For DW1, the bound states are localized and may have different energies for different bright spots. Along the step of the substrate, we don't see the bound state formation although the film must be somehow strained because of the existence of curvature across the step.

In conclusion, the 2-QL $Bi_2Te_3$ and $Sb_2Te_3$ films behave differently in the domain structures. There are two kinds of domain boundaries DW1 and DW2 having small and large rotational angles, respectively. While in 2-QL $Bi_2Te_3$ there is only DW1, there are both DW1 and DW2 in 2-QL $Sb_2Te_3$. By the spatially dependent STS, we show that DW1 and DW2 disturb the electronic structures of the ultra-thin films differently. Both DW1 and DW2 introduce bound states along the domain boundaries. The difference is that the bound states in DW1 is localized while those in DW2 is delocalized. The delocalization of bound states in DW2 might contribute to the transport in the ultra-thin films of this material. Our work points out that to realize the edge-state dominated transport at levitated temperatures, domain boundaries of type DW2 (if there exist) need to be suppressed. $Bi_2Te_3$ is not a good candidate of possible QSH states because of its buried Dirac point. For the ultra-thin $Sb_2Te_3$ film or its derivatives, further work must be done to minimize the formation of DW2 either by refining the growth parameters or using suitable substrates.

## Acknowledgments

We acknowledge the supporting from National Science Foundation (Grants No. 61804056, 92065102). We also acknowledge support from Open Research Fund Program of the State Key Laboratory of Low-Dimensional Quantum Physics, Tsinghua University.

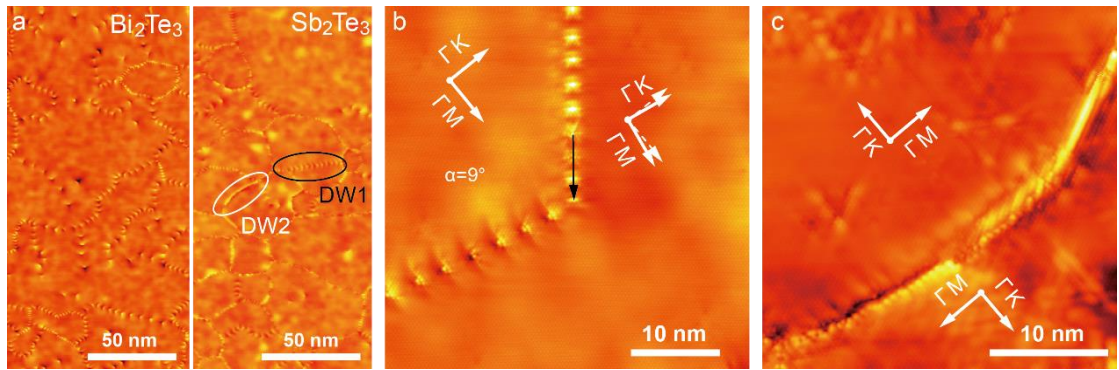

Figure 1. a. STM images (3.0 V, 10 pA) of 2-QL $Bi_2Te_3$ (left) and $Sb_2Te_3$ (right) films. DW1 and DW2 indicate two different kinds of domain boundaries. b and c. High-resolution STM images of domain boundaries of type DW1 and DW2, respectively. The $\overline{\Gamma}-\overline{M}$ directions of the films' crystal structure projected onto the surface brillouin zone are labeled.

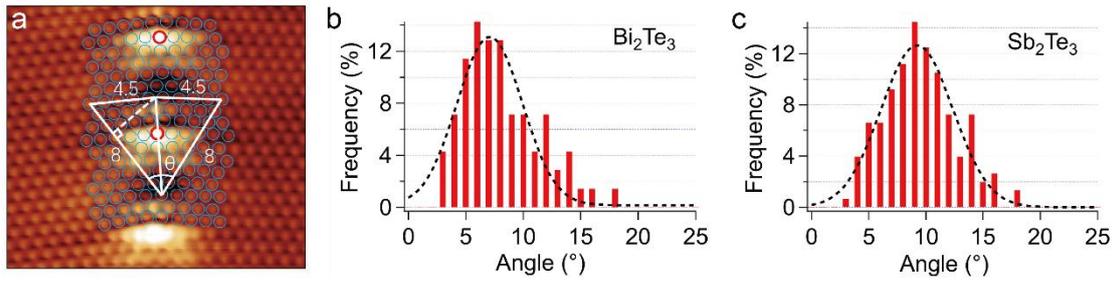

Figure 2. a. A Zoom-in STM image of figure 1b showing the typical atomic resolved DW1 domain boundary. b and c. Statistics of rotation angles between domains in the 2-QL $Bi_2Te_3$ and $Sb_2Te_3$ films. The red circles mark the position where one atom is inserted. These atoms have a higher apparent height, shown in the image as bright spots. The rotation angle between two adjacent domains $\alpha = \theta - 60°$.

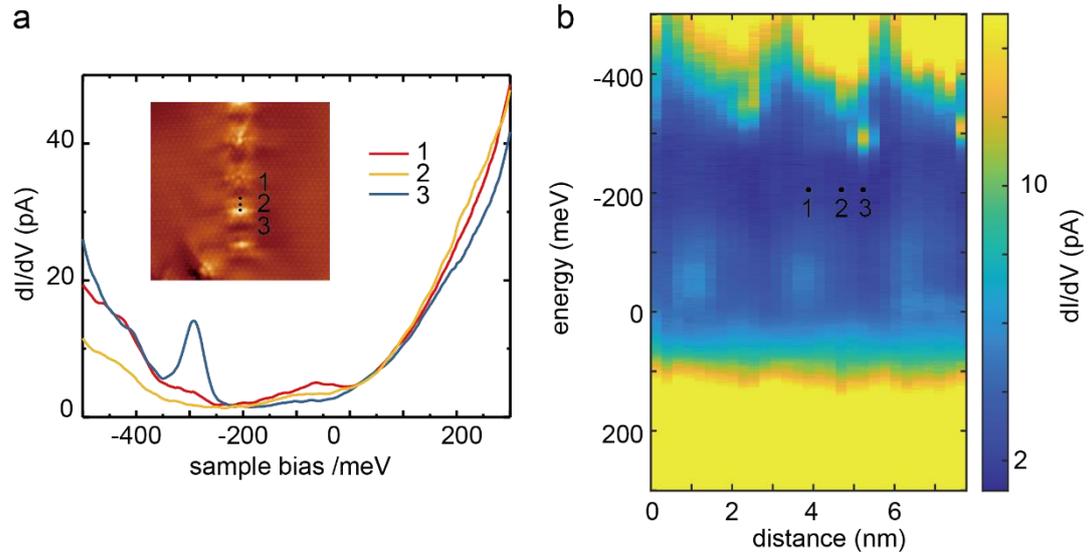

Figure 3. a. Typical STS along DW1. Type 1 is the STS between bright spots. Type 2 is the STS ~0.5 nm away from the bright spots. Type 3 is the STS on the bright spots. b. Spatial dependent STS along the arrowed line indicated in figure 1b. The spectra are taken step-wisely with an interval of ~0.25 nm and are plotted without interpolation.

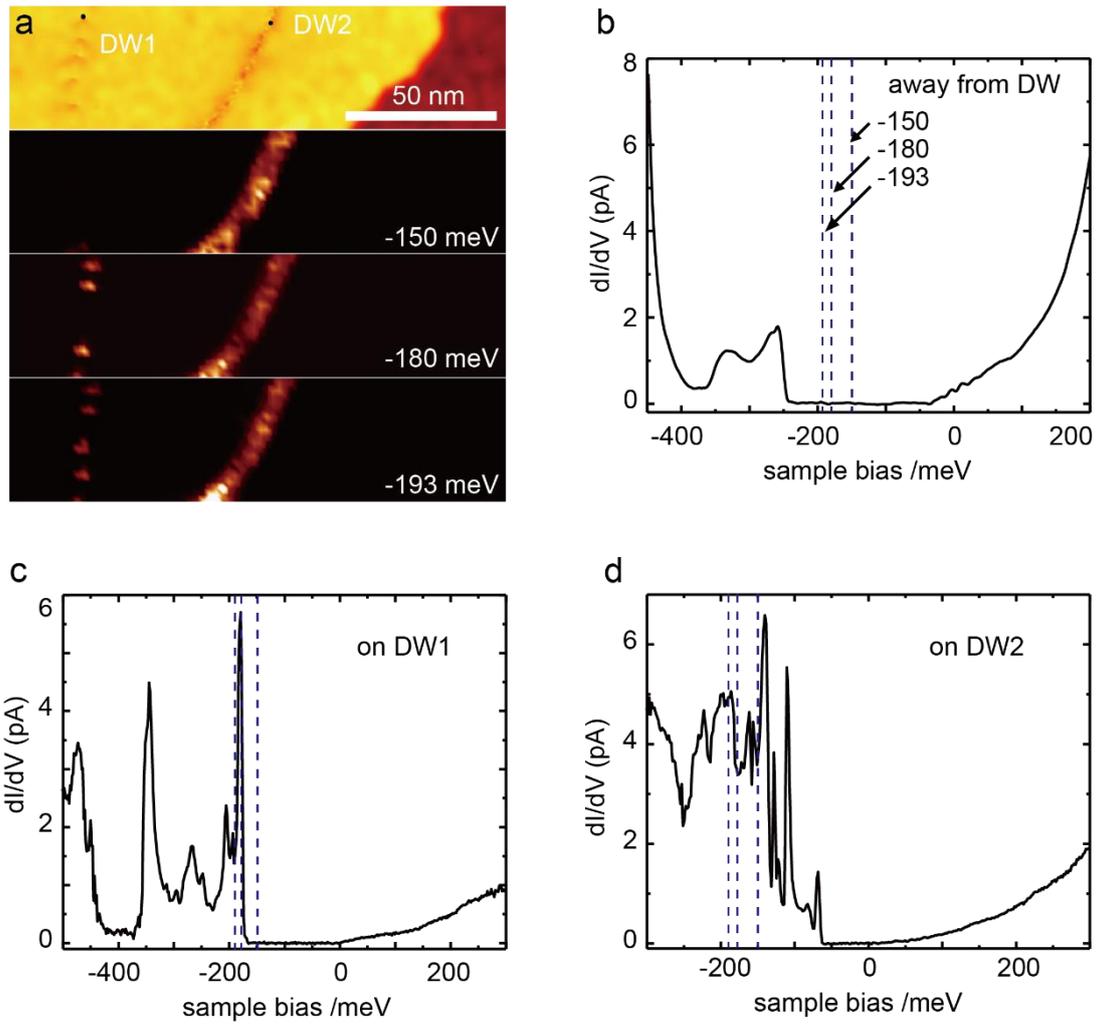

Figure 4. a. An STM topographic image and the corresponding STS mapping at various energies on the 2-QL $Sb_2Te_3$ film. In this region, DW1 and DW2 are indicated. b, c and d. STS taken at position away from domain boundaries, on the bright spot of DW1, on DW2 (black spots in a), respectively. The dashed lines indicate the energies at which STS mappings in a are taken.